\documentclass[preprint,proceedings]{rmaa}
\SetYear{2003}
\SetConfTitle{IAU Colloquium 194: Compact Binaries in Galaxy and Beyond}
\title{The Radio : X-ray Correlation in Cygnus X-3 and other Galactic Microquasars}
\author{M. Choudhury, \altaffilmark{1} A. R. Rao \altaffilmark{1}}
\altaffiltext{1}{Tata Institute of Fundmental Research, Mumbai-400005. India.(\protect\email{manojulu@tifr.res.in})}
\suppressfulladdresses
\listofauthors{M. Choudhury \& A. R. Rao}
\indexauthor{Choudhury, M.} 
\indexauthor{Rao, A. R.}

\addkeyword{Accretion}
\addkeyword{Accretion Disks}
\addkeyword{Radio Continuum: Stars}
\addkeyword{X-Rays: Binaries}

\begin{document}

\maketitle 

\boldabstract{Cygnus X-3, a Galactic X-ray binary, shows the presence of outflows in the form of radio jets. The SED in the X-ray band shows a complicated structure and evolution. We review some recent results of the long-term correlation of the radio and X-ray emissions, chiefly in the low (hard) states of the source. Comparing the results with those of other Galactic microquasars, we attempt to provide a consistent picture of the accretion -- ejection mechanism in these sources.}

\begin{figure}[!t]
\includegraphics[height=8cm,width=0.65\columnwidth,angle=-90]{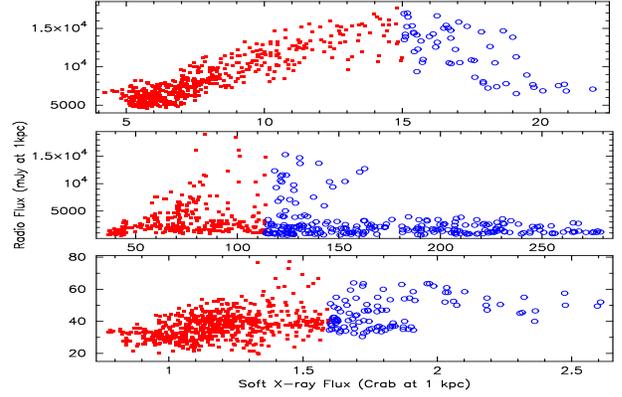}
\caption{Scatter diagram of soft X-ray emission and radio emission of Cygnus X-3 (top panel, GRS 1915+105 (middle panel) \& Cygnus X-1 (bottom panel), in the low as well as hard state, after removing the radio flares.}
\label{fig 1}
\end{figure}

\begin{figure}[!t]
\includegraphics[height=8cm,width=0.60\columnwidth,angle=-90]{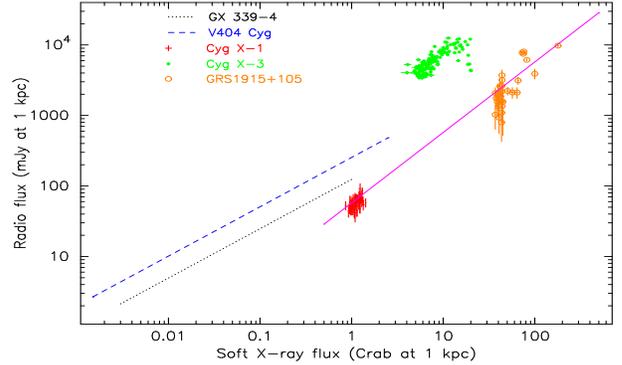}
\caption{Soft X-ray~:~radio correlation across various sources spanning 5 orders of magnitude of emitted flux.}
\label{fig 2}
\end{figure}

The X-ray binary Cygnus X-3 exhibits mainly two types of X-ray emissions, high
(correspondingly soft) and low (correspondingly hard), although the individual
spectral components are different compared to the canonical black hole states
(Choudhury \& Rao 2002).  During the low (hard) state, the soft X-ray emission
(2--12 keV) is very strongly correlated to the radio emission (2.2 GHz), whereas the
hard X-ray (20-100 keV) is anti-correlated to the both soft X-ray and radio
(Choudhury et al. 2002). This is due to the pivoting of the X-ray spectral energy
distribution (SED), correlated to the radio emission, with the pivot point lying
between 10 -- 20 keV. Similar behaviour of the pivoting of the X-ray SED with the
radio emission is seen in the persistant black hole candidates GRS 1915+105 and
Cygnus X-1 (Choudhury et al. 2003), with the pivot point of GRS 1915+105 lying near
about 30 keV, while that of Cygnus X-1 lies in the region of 50 -- 100 keV
(Zdziarski et al. 2002). In the high (correspondingly soft) state, the radio emission
shows a characteristic quenching of flux (not considering the flares) with the
increase in X-ray flux (Figure \ref{fig 1}). GX 339-4 also shows similar pivoting of
the X-ray spectra above 300 keV (Wardzinski et al. 2002) in the low-hard state,
correlated to radio emission, with quenched radio emission in the high-soft state
(Corbel et al. 2000). The uniform behaviour of the various sources spanning 5 orders
of magnitude of emitted flux (Figure \ref{fig 2}) suggest the presence of uniform
physical mechanism in this class of objects, viz. Two Component Accretion Flow (TCAF)
model of Chakrabarti (1996). The Comptonising region is confined within the CENBOL,
the extent of which, depending on the accretion rate, determines the X-ray SED
evolution, while the radio emission is a function of the compression ratio, showing
a turnover at the X-ray spectral transition.

\end{document}